\documentclass{epl}
\newcommand{\be}{\begin{equation}}
\newcommand{\ee}{\end{equation}}
\title {An $\epsilon$-expansion for Small-World Networks}
\author{M. B. Hastings}
\institute{
T-13 and Center for Nonlinear Studies, Los Alamos National
Laboratory, Los Alamos, NM 87545, hastings@lanl.gov 
}
\pacs{89.75.Hc}{Networks and genealogical trees}
\pacs{64.60.Ak}{Renormalization-group studies of phase transitions}
\date{July 1, 2004}
\begin{document}
\maketitle
\begin{abstract}
I construct a well-defined expansion in $\epsilon=2-d$ for
diffusion processes on small-world networks.  The technique permits
one to calculate the average over disorder of moments of the Green's
function, and is used to calculate the average Green's function and
fluctuations
to first non-leading order in $\epsilon$, giving results which agree with
numerics.  This technique is also applicable
to other problems of diffusion in random media.
\end{abstract}

The small-world network\cite{ws} has served as a fundamental model in the
field of networks\cite{net}.  However, the problem of averaging over
the possible different random connections in the small-world network is
severe in low dimensions: a study\cite{sw1} of the properties of even the
simple problem of diffusion on the one-dimensional network leads to a difficult
problem that, thus far, has only been tackled approximately.

The physical reason for this problem is a breakdown of mean-field 
theory\cite{khk} in dimensions $d$ less than two, and the emergence of
strong site-to-site fluctuations of the Green's function, so that
the properties of the system cannot be represented by simply studying
the average.  However, this opens the possibility of perturbing in
$\epsilon=2-d$, as will be shown in this paper.

The small-world network is constructed by starting with a regular
lattice in $d$-dimensions.  Then, some set of long-range links are
added: a given pair of sites $i,j$ is connected with probability
$pa^d/V$, where $V$ is the total number of sites in the system.  Here, we
define a length $a$ as the lattice scale, and $p$ as the {\it density} of links.
Then, as $V\rightarrow\infty$, each site has a Poisson distribution
of links emanating from it, with on average $pa^d$ links.
Typically, the links, if any, leaving a given site
will connect that  site to other sites far away in the system.

Looking for universal results,
we consider the case of a low density of links, $pa^d<<1$.
Ignoring sample-to-sample fluctuations,
the natural mean-field system to consider
is one in which each site is coupled to all others with a strength
$\sim p/V$.  This leads to a solvable problem
with a correlation length $\xi\propto p^{-1/2}$, or a correlation volume
$\xi^d\propto p^{-d/2}$.  Returning to the original problem with fixed links,
we see that such a volume has $p^{1-d/2}$ links in it, and as $p\rightarrow 0$,
this number of links tends to zero for $p\leq 2$.  This is a major problem.
Mean-field theory ignores fluctuations in the number of links, which
is only justified if the number of links is large, a condition which is not
satisfied in this case for $d<2$.  This problem is a result of a violation
of the modified Harris criterion introduced in \cite{mft}.

Instead, we expect that the correlation
length for the average Green's function 
must be at least $p^{-1/d}$ for small $p$,
so that there is on average at least one link in a correlation volume.  We
will see below that the correlation length is in fact proportional to
$p^{-1/d}$ in this limit,
and the $\epsilon$-expansion will enable us to calculate the
prefactor, as well as to study fluctuations about the average.
Given a $p^{-1/d}$ scaling of the correlation length,
the number of links in a correlation volume is some $p$-independent
number.
The basis for the $\epsilon$-expansion is the observation that, within
a self-consistent approximation, this number diverges as $1/\epsilon$,
so that fluctuations about a self-consistent mean-field
can be calculated diagrammatically as done below.

{\it Green's Function---}
Our general problem is to study diffusion on the small-world network.
Thus, we must calculate Green's functions of the following
problem:
$\partial_t \rho_i(t)=-\sum_j \Gamma_{ij} \rho_j$,
where $\rho_i(t)$ is the probability of finding some randomly walking particle
at site $i$
and $\Gamma$ is the Laplacian on the small-world network.  We take
\be
\label{geq}
\Gamma=\Gamma^0+q U.
\ee
Here, $\Gamma^0_{ij}$ is the Laplacian on the regular part of the network.
$\Gamma^0_{ij}=-a^{-2}$ if $i$ and $j$ are neighboring sites on the regular
network, while $\Gamma^0_{ii}$ is equal to $a^{-2}$ times
the coordination number of site $i$.  $U_{ij}$ is the Laplacian on the
long-range links.  $U_{ij}=-1$ if $i,j$ are connected by a long-range
link, where $U_{ii}$ is equal to the number of long-range links leaving
site $i$.

We have inserted an extra factor of $q$ multiplying the matrix $U$ in
Eq.~(\ref{geq}).  If $q$ is small and $p$ is large, this implies that
we have a high density of weak links and the problem can be solved via
a mean-field theory in which we ignore the fluctuations in the local
density of links.  However, we will be interested in the opposite case, where
$q$ is of order $1$, while $p$ is small.

Fourier transforming, we are interested in the Green's function
$(i\omega+\Gamma)^{-1}$.  
However, we will focus on the case 
$\omega\rightarrow 0$.
The matrix
$\Gamma$ has a zero single mode, due to the conservation of $\sum_i \rho_i$ by
the diffusion process.
Throughout, we will work in the subspace orthogonal
to this zero mode, defining
$G=\lim_{\omega\rightarrow 0}(i\omega+\Gamma)^{-1}$ and 
$G^0=\lim_{\omega\rightarrow 0}(i\omega+\Gamma^0)^{-1}$ in this subspace.  
This Green's
function is related to the return probability of a random walker, and
also to the roughness
of a surface defined by Edwards-Wilkinson dynamics on the network\cite{khk}.
We will
compute
$\overline{G_{ij}}$, where the line denotes averaging
over the ensemble of different random networks, as well as computing
higher moments such as
$\overline{G_{ij}G_{kl}}$.

For $p$ small (compared to $a^{-d}$),
the length $p^{-1/d}$ is much larger than the lattice scale, so that
we can take a continuum limit, setting
$\Gamma^0$ equal to the continuum Laplacian $\partial^2$.
In the limit of small $p$, the probability of a single site having
more than one link becomes vanishingly small.  This continuum limit
leads to ultraviolet divergences for $d\geq 2$, which are cutoff
at the lattice scale $a$.  However, for $d<2$, this
limit is completely convergent, and thus for $d<2$ we find universal
results, independent of the lattice details.  This use of a continuum limit
is essential to continue the results to arbitrary real dimension $d<2$.

For clearer notation, in the continuum limit we will often use $d$-dimensional
vectors $x,y,...$ to label lattice sites, rather than indices 
$i,j,...$.
We write $G(x,y)$ to
denote a Green's function between points $x,y$ and $\overline{G(x)}$ to denote
a Green's function $\overline{G(x,0)}$
(the averaging restores translational symmetry).
We also use a vector $k$ to label momenta, defining
the averaged Green's function
at Fourier mode $k$ to be $\overline{G(k)}=\int {\rm d}^d x \overline{G(x)}
\exp(i k \cdot x)$.  We use a similar notation for
other matrices in the continuum limit.  Discrete $\delta$-functions
get replaced by Dirac $\delta$-functions, while sums get replaced by integrals.

Ignoring fluctuations in the local density of links,
the mean-field approximation consists of replacing $U$ by $\overline U$,
so in the continuum limit
$\overline{G}\approx(\partial^2+\overline{U})^{-1}$.  Then, we find that
$\overline{U(x,y)}=pq\delta(x-y)-pq/V$.  
This gives a Green's function
$\overline {G(x)}=\int{\rm d}^dk(2\pi)^{-d} \exp(i k \cdot x)
(k^2+pq)^{-1}$.  The correlation
length is $(pq)^{-1/2}$ as discussed above, and thus this expansion must
breakdown for $d<2$ in the limit of small $p$ at fixed $q$.

{\it Self-consistent Calculation---}
To go beyond the mean-field calculation, we use impurity-averaged
perturbation theory\cite{iapt}, following \cite{khk}.  We continue
with lattice notation here, since this development is not
specific to the continuum limit.
The perturbative expansion of $\overline G$ in powers of $U$ is
$\overline G=G^0-\overline{G^0 U G^0}+\overline{G^0 U G^0 U G^0}-...$,
where a product of matrices is implied.
Each of the terms in this expansion can be computed using the given distribution
of disorder.  The mean-field calculation is based on the following
approximation: $G\approx 
(\Gamma^0+\overline U)^{-1}=G^0-G^0\overline U G^0 + G^0 \overline U
G^0 \overline U G^0 -...$  These two expansions are equal at
zeroth and first order in $U$, but differ at second order in $U$, by
an amount depending on the second cumulant of $U$.
\begin{figure}
\onefigure[scale=0.5]{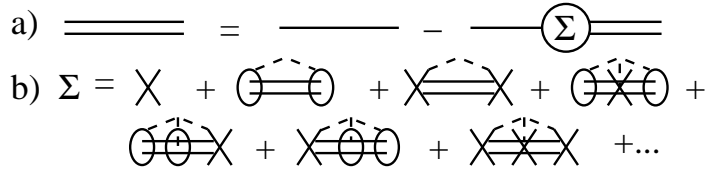}
\caption{a) Diagrammatics for $\Sigma$ and b) Self-consistent sum of
interactions with a single link.}
\label{fig1}
\end{figure}

It is useful to introduce a diagrammatic notation for the perturbative
calculations, as shown in Fig.~1.  We use a single solid line to denote
$G^0$, and a double solid line to denote $\overline G$.  We use
a cross with no dashed lines attached to denote the average $\overline U_{ii}$.
A pair of crosses connected by a dashed line is used to denote an
average $\overline{U_{ii}U_{jj}}-\overline{U_{ii}}\;\overline{U_{jj}}$.
Three or more crosses connected by dashed lines are used to denote the
third and higher cumulants.  These diagrams denote two, three, or more
scatterings of a single link.  Circles are used to denote averages of
off-diagonal terms, $U_{ij}$ for $i\neq j$.  Dashed lines can also connected
both circles and crosses, again denoting higher cumulants.
Within this diagrammatic approximation, the weight of a given diagram is
equal to $p^{n_l} (-q)^{n_s}$.  Here $n_l$ is
equal to the number of sets of
circles or crosses connected by dashed lines, while $n_s$ is equal to the
total number of circles plus crosses.  Thus, each link gives one factor of $p$
(the probability of finding a link) multiplied by $-q$ to the number of times
that link appears in $U$.

We note that
a single circle, not connected by a dashed line, denotes $\overline{U_{ij}}$,
which vanishes as $V\rightarrow \infty$, and thus may be
ignored in this perturbation expansion.  In general, any diagram
involving an odd number of off-diagonal terms for a given link vanishes
as $V\rightarrow\infty$.

Let us return to the continuum limit for specific calculations.
We introduce the self-energy $\Sigma(k)$ by setting $\overline {G(k)}=
[\Gamma^0+\Sigma(k)]^{-1}$, as
shown diagrammatically in Fig.~1(a).
Approximating $\Sigma$ to leading order in $U$ we get
$\Sigma(k)=\overline {U(k)}$.  We have $\Sigma(0)=0$,
However, for all $k\neq 0$, $\Sigma(k)$ is $k$-independent,
so that, outside of the subspace of the zero mode,
we may write $\Sigma(k)=\Sigma=pq$.
We now compute corrections to this
result, to second order in $U$.  To this order, $\Sigma(k)$ is still
$k$-independent for $k\neq 0$.  This self-consistent Born approximation
is defined by the first three diagrams on the right-hand side
of Fig.~1(b).  Higher
diagrams of Fig.~1(b) include all diagrams with arbitrary numbers
of scatterings off a {\it single} link.  We
self-consistently
use double lines for the Green's function in Fig.~1(b),
giving $\Sigma=pq-
2pq^2 \overline{G(0)}$, where
\be
\label{g0eq}
\overline{G(0)}=\int \frac{{\rm d}^d k}{(2\pi)^d} \frac{1}{k^2+\Sigma}.
\ee
For $d<2$, the integral of Eq.~(\ref{g0eq}) is convergent, and equal to
\be
\label{gfgf}
\frac{\pi^{d/2}}{(2\pi)^d} \Sigma^{d/2-1} \Gamma(1-d/2).
\ee
The self-consistent Born approximation at second order in $U$ then becomes
$\Sigma=pq-2 pq^2 \overline{G(0)}$.  As a first guess at solving this
equation, we substitute the first order result for $\Sigma$ in the
equation for $\overline{G(0)}$, getting 
$\Sigma=pq-2 p^{d/2}q^{d/2+1} \pi^{d/2}\Gamma(1-d/2)/(2\pi)^{d}$ for $d<2$.
The second order correction is comparable to the first when
$q\sim p^{2/d-1}$.  Thus, for
$q<<p^{2/d-1}$, perturbation theory may be applied.  We will be interested
in the opposite limit of $p$ small at fixed $q$ so that
$q>>p^{2/d-1}$ and perturbation theory breaks down for $d<2$. 

A better approximation for $\Sigma$ is to include all diagrams in
Fig.~1(b), summing all diagrams involving interactions with
a single link.
Since we are interested in the case in which
the density of links is small, but the scattering $q$ off a single link
is of order unity, we sum all scatterings off a single link.
This gives
$\Sigma=pq/[1+2 q \overline{G(0)}]$.
For $p$ small, we will find $\overline{G(0)}>>1$, so that we can take
$\Sigma=pq/[2q\overline{G(0)}]=p/[2\overline{G(0)}]$.  
Here, the factor of two arises from the different possible diagonal
and off-diagonal scatterings
off a single link: if a particle interacts with a given link $n$ times,
then there are $2^{n-1}$ total diagrams contributing to $\Sigma$.
Using 
Eq.~(\ref{gfgf}) for $\overline{G(0)}$, we find for $d<2$ that
the solution of 
$\Sigma=p/[2\overline{G(0)}]$ is $\Sigma=$
\be
\label{scp}
\Sigma_0=[(2\pi)^d p/(2 \pi^{d/2}\Gamma(1-d/2)]^{2/d}.
\ee
We have placed the subscript $0$ on $\Sigma$, since later this result
will be used as a zero order approximation, with corrections to it
in powers of $\epsilon$.
This gives $\Sigma\propto p^{2/d}$, so that the correlation length
varies as $p^{1/d}$.  We will find below that a more careful treatment
leads to universal corrections
in orders of $\epsilon$
to the prefactor of Eq.~(\ref{scp}), but does not change the scaling with
$p$.

In contrast,
for $d>2$, the integral of Eq.~(\ref{g0eq}) is divergent at large $k$, but
it converges at small $k$ even for $\Sigma=0$.  The divergence at large $k$
is cut off by the lattice scale, $a$.  Thus, $\overline{G(0)}$ has a
well-defined limit as $\Sigma\rightarrow 0$; this limit has a non-universal
dependence on the lattice details and is equal to some number, $g$.
The self-consistent equation
becomes $\Sigma=pq/[1+2 q \overline{G(0)}]$.  To leading order
in $p$, the solution of this equation gives
$\Sigma=pq/(1+2 q g)$.
The magnitude of the corrections
to the mean-field result, $\Sigma=pq$,
depends on the product $qg$ and does not have any universal behavior.  There
are also further corrections to the mean-field result which involve scattering
off multiple links.  These corrections will not be considered here.

Finally, consider $d=2$.  In this case, we will show below that the
approximation above leads to {\it exact} results for the small $p$ behavior of
$\Sigma$.  The integral of Eq.~(\ref{g0eq}) is logarithmically divergent.
The divergence is cut off at a $k$ of order $a^{-1}$, giving
$\overline{G(0)}=-(2\pi)^{-1}\log(a\Sigma^{1/2})$.  The self-consistent
equation then gives
$\Sigma=pq/[1+2q\overline{G(0)}]$.  As for $d<2$, $G(0)>>1$ for small
$p$ so that this reduces to
$\Sigma=p/[2\overline{G(0)}]=-2\pi p/\log(a^2\Sigma)$.  
Solving this self-consistent equation for $a^2p<<1$ gives
$\Sigma=-2\pi p/\log(pa^2)$, plus subleading terms.

{\it Corrections in Two Dimensions and Renormalization Group---}
Thus far, the calculation has followed \cite{khk}.  Now, however,
we go beyond the self-consistent calculation to obtain an
expansion in $\epsilon$.  The first step is to analyze the behavior
for $d=2$ in more detail, and show that the results above
are exact for the leading scaling of $\Sigma$ with $p$.  In the next
section, a diagrammatic perturbation expansion will be presented to
compute results in powers of $\epsilon$.

In $d=2$, the correlation volume is of order $1/\Sigma$.
The physical motivation for the results below
is that for $\Sigma\sim -p/\log(pa^2)$, the average
number of links in a correlation volume, $p/\Sigma$,
is of order $-\log(pa^2)$.  Thus, the average
number of links in a correlation volume diverges as $p\rightarrow 0$.
This, however, is exactly the condition we need to make mean-field theory
work, as it enables us to ignore fluctuations in the number of links.

Turning away from perturbation theory, we now consider the problem
scale-by-scale, starting with the shortest distances.  We begin
with a physical description of the problem, followed by a more
careful calculation below.
On length scales $l<<p^{-1/d}$,
each link can be considered independently, since
the density of links at such scales is very small: the probability of
finding a link near any other link is negligible.
Solving the problem
of a single link at a scale $l$ is very similar to the calculation done above.
We must sum multiple scatterings off the single link.  This will renormalize
the interaction strength of the link from $q$ to
$q/[1+2q G_l(0)]$, where $G_l(0)\sim (2\pi)^{-1}\log(l/a)$ is the 
Green's function of matrix $\Gamma^0$ at
scale $l$.  
Any function which cuts off momenta less than $l^{-1}$ will suffice for
$G_l$.  We choose $G_l(k)=(k^2+l^{-2})^{-1}$.
Then, once we reach scale $l=p^{-1/2}$, the link interaction
strength has been renormalized to 
$q_{ren}=q/[1+2 q G_{l}(0)]\sim -2\pi/\log(pa^2)$.

The number of links in an area of size $a^2$ was $pa^2$.  In contrast, on the 
new scale $l=p^{-1/2}$, the number of links in area $l^2$ is of order unity,
while the scattering strength $q_{ren}$
is small, and thus corrections to mean-field theory become negligible
for small $p$.  Then, we find $\Sigma=pq_{ren}=-2\pi p/\log(pa^2)$.

{\it Expansion in $\epsilon$---}
Now, we wish to compute quantities for $d<2$.  In this case, we find that
once we reach a scale $l\sim p^{-1/d}$, the link interaction
strength $q_{ren}$ has been reduced to an amount of order
$q/[1+2 q G_{l}(0)]\sim \epsilon l^{-\epsilon}$.  Unlike the
case in two dimensions, $q_{ren}$ does not vanish as we take the limit
$pa^d\rightarrow 0$.  However, we will find that
the interaction strength of the links is of order $\epsilon$, so that
corrections to mean-field theory can be
computed in powers of $\epsilon$.
The $\epsilon$-expansion is defined by a simple re-ordering of
diagrams.  First, note that Eq.~(\ref{scp}) defines a length scale $l$ by
\be
\label{leq}
l=\Sigma_0^{-1/2}=[(2\pi)^d p/(2 \pi^{d/2}\Gamma(1-d/2)]^{-1/d}.
\ee
so $G_l(k)=(k^2+\Sigma)^{-1}$.  Then, we define the
resummed diagonal scattering off a single link as in Fig.~2(a), denoted by
a cross
with double dashed lines.  Here, the
single line denotes the Green's function $G_l$, rather than $G^0$ as before
(this resummation is
closely related to the $T$-matrix in scattering theory).
The resummed off-diagonal scattering is defined similarly by
a circle with double dashed lines.

\begin{figure}
\onefigure[scale=0.5]{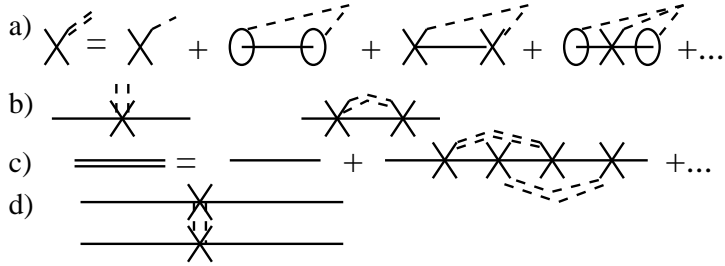}
\caption{a) Definition of resummed scattering.  b)  Disallowed diagrams.
c) Leading corrections to ${\overline G}$.  d)  Leading contribution
to $\overline{G G}-\overline G \, \overline G$.  For all diagrams, single
line denotes $G^l$ in this figure.}
\label{fig2}
\end{figure}
The starting point for the $\epsilon$-expansion
is the self-consistent calculation of Eq.~(\ref{scp}).
That expansion resums a large set of diagrams known as the
``rainbow" diagrams; this is the set of all diagrams except those in
which a particle first scatters off one link, then off a second, then
returns to the first link, and then returns to the second link (possibly with
additional scatterings off other links included).  We now present an
expansion for $\overline G$
which exactly includes all of the missing diagrams, such as in Fig.~2(c),
with the correct coefficent, using resummed scatterings with each link.  We
then show that this
leads to an expansion in powers of $\epsilon$.

First, write down all possible diagrams, using $G_l$
for the Green's functions and double dashed lines for interactions
with links, subject to the following two constraints: $(1)$ double dashed
lines must always connect at least two crosses or circles.  That is, single
crosses or circles never appear individually, so the first diagram of Fig.~2(b)
is not allowed.  This constraint on diagrams is included because
such diagrams are already taken into account in the self-consistent expansion,
by the replacement of $G^0$ by $G_l$.
$(2)$ two successive interactions with links must always
involve different links (links not connected by dashed lines),
so that the second diagram of
Fig.~2(b) is also not allowed.  The second constraint is presented to avoid
overcounting multiple scatterings.
Given these constraints, the leading
correction to the averaged Green's function is given by Fig.~2(c).
 Each diagram is assigned a factor of $p^{n_l}(-q_{ren})^{n_s}$, where
$n_l$ is now the number of sets of circles and crosses connected by
double dashed lines and $n_s$ is again the total number of circles and crosses.
Here, $q_{ren}=q/[1+2qG_l(0)]=\Sigma_0/p.$

This expansion, though it simply reorders the diagrams, perfectly realizes
the desired treatment of the problem scale-by-scale.  We must start with
the shortest scales.  However, at the shortest scales, the only processes
involve multiple scattering off single links; only when the scale becomes
of order $p^{-1/d}$ do processes with multiple links become important.  Thus,
we resum scattering off the single link, up to the scale $l$.  Unlike the
case for $d=2$, processes involving multiple links, such as that
shown in Fig.~2(c), are no longer vanishing, but we now show that
they are higher
order in $\epsilon$.
First, it is convenient to rescale
all distances by $l$.  Then, we set ${\tilde x}=x/l$, 
${\tilde k} =k l$.  We use the Green's function
$G_1(k)=(k^2+1)^{-1}$, and define $\tilde p=pl^d$, while
$\tilde q_{ren}=q_{ren} l^{2-d}$.  Then, we define 
$\overline {\tilde G(\tilde x)}$ by
$\overline {G(x)}=\overline{\tilde G(\tilde x)} l^{2-d}$.  Then,
the perturbation expansion for $\overline{\tilde G}$ is
obtained by using the same set of diagrams as above, but replacing
$p$ by $\tilde p$, $q_{ren}$ by $\tilde q_{ren}$, and $G_l$ by $G_1$.
Now, we have
\begin{eqnarray}
\tilde p=2 \pi^{d/2}\Gamma(1-d/2)/(2\pi)^d=\epsilon^{-1}4\pi^{d/2}/(2\pi)^d+...,
\\ \nonumber
\tilde q_{ren}=\tilde p^{-1}=\epsilon (2\pi)^{d}/(4\pi^{d/2})+...,
\end{eqnarray}
where here we have given a series expansion of the results for 
$\tilde p, \tilde q_{ren}$,
and the $...$ denote higher order terms in $\epsilon$.
Thus, we have defined a new problem of scattering with links of strength
$\epsilon$ and density $\epsilon^{-1}$; this a problem of a high density
of weak links so that perturbative techniques work well and lead to
an expansion in $\epsilon$.  
Physically, one can imagine that for $d$ close to two, a randomly diffusing
particle has only a weak interaction with a link: the dimension of the
path of the particle is two, so that for $d$ close to two the particle
can easily ``miss" a given link.

There are
a finite number of diagrams at each order in $\epsilon$ as we now
show.  The number of
scatterings of each links is at least two by the rules above, so
$n_s\geq 2n_l$.  Then, $\tilde p^{n_l} \tilde q_{ren}^{n_s}\sim
\epsilon^{n_s-n_l}$ is at least order $\epsilon^{n_l}$.  Thus, to
order $\epsilon^n$, we need only consider diagrams $n_l\leq n$.
For a given $n_l$, we need only consider diagrams with $n_s\leq n+n_l$, leaving
us with only a finite number of diagrams at each order in $\epsilon$.
It is important to consider the possibility of ultraviolet divergences
in two dimensions, since they may give extra factors of $\epsilon^{-1}$.
However, the only ultraviolet divergences in two dimensions in this
expansion arise from
Green's functions $G_1(0)$ and this expansion resums all
such divergent diagrams which involve scattering off at least one link:
the only divergent contribution to $\overline{G(0)}$ is
from the very first diagram in Fig.~2(c).  Finally, to all orders
in $\epsilon$, the scaling of $\overline{G(0)}$ with $p$ is unchanged
from the self-consistent calculation.

We now use this formalism to compute specific results.  We first
compute the average Green's function.  From Fig.~2(c), we have
$\overline{G(0)}=l^{2-d}[G_1(0)+\tilde p^2 \tilde
q_{ren}^4 \int \int {\rm d}^dx {\rm d}^d y G_1(x) G_1(y) G_1(x-y)^3+...]$.
We have numerically evaluated this integral in $d=2$ to get the leading
corrections in $\epsilon$ (to simplify this integral, we used the trick
that it equals $-(1/4)
\partial_l \int {\rm d}^2x G_l(x)^4$ at $l=1$).  The result is
$\overline{G(0)}=l^{2-d}[G_1(0)+\epsilon^2 \pi^2 (.00106...)+...]$,
where we have also evaluated $\tilde p,\tilde q_{ren}$ for $d=2$, and
where $l$ is given by Eq.~(\ref{leq}).  Here, $G_1(0)=\pi^{d/2}/(2\pi)^d
\Gamma(1-d/2)$ and hence diverges as $\epsilon^{-1}$.  As noted above,
this is the only ultraviolet divergent diagram.

To go to higher order in $\epsilon$, it would be necessary to keep
additional diagrams, as well as to expand the integrals in powers
of $\epsilon$ near $d=2$.  Since we are interested in $d=1$, let us
first evaluate Fig.~2(c) in $d=1$, where
$\tilde p=\tilde q_{ren}=1$, so
\be
\label{2c}
\overline{G(0)}=p^{-1}[G_1(0)+\int
\int {\rm d}x {\rm d}y G_1(x) G_1(y) G_1(x-y)^3+...]=
p^{-1}(1/2+5/256+...).
\ee
At higher order, if we evaluate all diagrams in $d=1$,
but use the $\epsilon$ expansion to define the ordering of diagrams for us,
all the diagrammatic integrals can be performed exactly,
since $G_1(x)$ then has a simple exponential decay.
This is a task left for future work, as is a test of
the convergence properties of this expansion.
Eq.~(\ref{2c}) compares well with numerics\cite{khk}.  There, $\overline{G(0)}$
was found to scale $p^{-1}$, with a prefactor slightly larger than $1/2$,
close to the result here; comparison of the exact difference between the
prefactor and $1/2$ would require higher order calculations here and
larger system sizes in the numerics.

Fluctuations in the Green's function can also be computed using
these techniques.  From Fig.~2(d), we have
$\overline{G(0)^2}-(\overline{G(0)})^2=
p^{-2} (\int{\rm d}x G_1(x)^4+...)=p^{-2}(1/32+...)$

{\it Discussion---}
We have presented an $\epsilon$-expansion for the properties of the
small-world network in $d=2-\epsilon$ dimensions, enabling us to compute
averages of moments of the Green's function.  This technique should, however,
have much greater generality.  In \cite{mft}, general criteria were
put forth for the breakdown of mean-field theory for any statistical
mechanical system
on a small-world network, in analogy to the usual Harris criterion for
disordered systems\cite{hc}.
Hopefully, for other systems in which these criteria are violated,
it will be possible to provide an $\epsilon$ expansion near
the critical dimensionality, analogous
to what has been done for the violation of the  ordinary Harris
criterion\cite{potts}.

These criteria can be extended to other problems than
networks, such as a randomly diffusing particle in an
array of traps\cite{rt}, a model inspired by work in reaction-diffusion
processes\cite{diff}.  The criteria\cite{mft} correctly predict
$d=2$ as the region for the breakdown of mean-field theory, and the
techniques here can be used to 
compute average properties
of the Green's function in $d=2$. 
For the random trap system
in $d=1$ with $p<<q$,
the particle has only a small probability to pass through
one trap to reach the next trap which is typically
far away.  This makes the trap system very far from mean-field theory and
suggests that the $\epsilon$-expansion will not converge to $\epsilon=1$;
however, it also enables the solution of the $d=1$ system by
considering each interval between traps
independently.  In contrast,
for the network, the particle can interact with
several links, even for
$p\rightarrow 0$, thus suggesting that the perturbative $\epsilon$-expansion
may work for the small-world case.
A true test will require
evaluation of additional higher order diagrams and is left for future work.

{\it Acknowledgements---} I thank G. Korniss, B. Kozma, and Z. Toroczkai for
useful discussions.  I thank the ICTP in Trieste, Italy for hospitality while
some of this work was completed.  This work was supported by DOE W-7405-ENG-36.

\end{document}